\documentclass[aps,prd,floats,floatfix,nofootinbib,twocolumn,10pt]{revtex4-1}
\usepackage{graphicx}
\usepackage{amssymb}
\usepackage{hyperref}
\usepackage{amsmath}

\begin{document}

\title{Constraining the Higgs couplings to up and down quarks using production kinematics at the CERN Large Hadron Collider}

\author{Gage Bonner}

\author{Heather E.~Logan}
\email{logan@physics.carleton.ca} 

\affiliation{Ottawa-Carleton Institute for Physics, Carleton University, 1125 Colonel By Drive, Ottawa, Ontario K1S 5B6, Canada}

\date{August 15, 2016}                                  

\begin{abstract}
We study the prospects for constraining the Higgs boson's couplings to up and down quarks using kinematic distributions in Higgs production at the CERN Large Hadron Collider.  We find that the Higgs $p_T$ distribution can be used to constrain these couplings with precision competitive to other proposed techniques.  With 3000~fb$^{-1}$ of data at 13~TeV in the four-lepton decay channel, we find $-0.73 \lesssim \bar{\kappa}_u \lesssim 0.33$ and $-0.88 \lesssim \bar{\kappa}_d \lesssim 0.32$, where $\bar{\kappa}_q = (m_q/m_b) \kappa_q$ is a scaling factor that modifies the $q$ quark Yukawa coupling relative to the Standard Model bottom quark Yukawa coupling. The sensitivity may be improved by including additional Higgs decay channels.
\end{abstract}

\maketitle 

\section{Introduction}
\label{sec:intro1}

If the Standard Model (SM) of particle physics is to be complete, then there must be a mechanism through which elementary particles acquire mass. The Higgs mechanism achieves this purpose, and a particle with the required properties was recently observed by the ATLAS and CMS collaborations at the CERN Large Hadron Collider (LHC)~\cite{Aad:2012tfa, Chatrchyan:2012xdj}. We can test some of the predictions of the SM by studying the Higgs boson's couplings to other particles. The SM does not numerically predict these couplings directly; it postulates relatively simple expressions for their size in terms of other observables. Therefore, if we can measure these observables (particle masses, mixing angles, etc.) while characterizing the strength of the Higgs couplings via its production and decay rates, we can determine whether or not the relations predicted by the SM are correct. This gives us a clue as to whether or not the SM Higgs mechanism actually provides masses for all constituents of the SM. 

For heavy gauge bosons $W$ and $Z$ we expect the Higgs couplings to be equal to $2 m_{W, Z}^2/v$ in the SM, where $v \approx 246\,\text{GeV}$ is the Higgs vacuum expectation value. The measured couplings have been found to be consistent with the SM within experimental error~\cite{Khachatryan:2014jba, Aad:2015gba}. In the fermion sector, we expect the Higgs couplings to quarks $q$ to be equal to $m_{q}/v$ in the SM. This is also true for charged leptons. The quantities $m_{q}/v$ are usually called the Yukawa couplings $y_q$. Since the couplings are proportional to the quark masses, we expect Higgs-mediated processes to be dominated by the heavy (top and bottom) quark contributions. Indeed, Higgs production is controlled mainly by gluon fusion whereby two gluons initiate a heavy quark loop which ejects a Higgs boson. There are several experimental analyses which probe the Higgs couplings to the heavy (top and bottom) quarks~\cite{Aad:2015gra, Aad:2014lma, Aad:2014xzb, Khachatryan:2015ila, Khachatryan:2014qaa, Chatrchyan:2013zna, CMS:2014zqa, Chatrchyan:2014vua}. These are also found to be consistent within uncertainties with the SM prediction, so we conclude that the SM Higgs mechanism is a valid theory for the origin of the heavy gauge bosons' and quarks' masses.

The situation is less clear for lighter quarks. Constraining the light quark Yukawa couplings is important since there are alternate models in which they differ from the SM expectation~\cite{Giudice:2008uua, Botella:2016krk, Harnik:2012pb, Bauer:2015kzy} or do not enter at all~\cite{Ghosh:2015gpa}. Constraints can be placed on the charm and strange quark Yukawa couplings using inclusive Higgs production rates in various SM decay channels~\cite{Meng:2012uj, Delaunay:2013pja, Perez:2015aoa, Perez:2015lra} and through exclusive radiative mesonic decays, $h \to V \gamma$, where $V$ is a charmonium or $s \bar s$ meson~\cite{Bodwin:2013gca, Kagan:2014ila, Koenig:2015pha} (see also Refs.~\cite{Delaunay:2013pja, Perez:2015lra}).  The charm Yukawa coupling is expected to be measured at a future International Linear $e^+e^-$ Collider to high precision using the anticipated excellent charm tagging in the low-background $e^+e^-$ collision environment~\cite{Baer:2013cma}.

Up and down quark Yukawa couplings are by far the hardest to constrain: at the LHC it is basically impossible to distinguish Higgs decays to up and down quark jets from $h \to gg$ or $h \to s \bar s$.\footnote{On the other hand, Ref.~\cite{Rentala:2013uaa} showed that a statistical discrimination between gluon jets and light-quark jets is possible using jet energy profiles.} Furthermore, since the cross section for quark fusion, $q \bar{q} \to h$, is proportional to the square of the relevant quark Yukawa coupling $y_{q}^2$, for SM couplings proton collisions are much more likely to result in $b \bar{b} \to h$ than $u \bar{u} \to h$ and $d \bar{d} \to h$ even though $u$, $d$ are the valence quarks of the proton.  In particular, the up and down quark masses are $m_u = 2.3^{+0.7}_{-0.5}~{\rm MeV}$ and $m_d = 4.8^{+0.5}_{-0.3}~{\rm MeV}$ ($\overline{\rm MS}$ masses evaluated at $\mu \simeq 2~{\rm GeV}$) while $m_b = 4.18 \pm 0.03~{\rm GeV}$ ($\overline{\rm MS}$ mass evaluated at $m_b$)~\cite{Agashe:2014kda}.  

It is customary to parametrize the deviations of the Yukawa couplings from their SM values using scaling factors $\kappa_q$~\cite{LHCHiggsCrossSectionWorkingGroup:2012nn}, so that the coupling terms in the Lagrangian become $-\kappa_{q} y_{q}^{\rm SM} \bar{q} q h$, with $\kappa_q = 1$ corresponding to the SM. We will adopt the convention of Ref.~\cite{Kagan:2014ila} in which the light quark couplings are all scaled relative to the bottom quark coupling. This greatly reduces the theoretical uncertainty in the reference coupling since the bottom quark mass has a much smaller experimental uncertainty than the up and down quark masses.  It also facilitates comparisons with the literature. Since the Yukawa couplings are proportional to the relevant quark mass, we have
\begin{equation} 
	\bar \kappa_q = \frac{m_q}{m_b} \kappa_q,
    \label{eq:kappa_bar_def}
\end{equation}
where $\bar{\kappa}_q$ is the light quark coupling scaled relative to that of the bottom quark. In the SM we expect $\bar \kappa_u \simeq 4.7 \times 10^{-4}$ and $\bar \kappa_d \simeq 1.0 \times 10^{-3}$~\cite{Kagan:2014ila}.

The current tightest constraints on up and down quark Yukawa couplings come from Higgs production and decay rates.  A global fit to all on-resonance Higgs data, allowing all of the Higgs couplings to vary, yields $|\bar \kappa_u| < 1.3$ and $|\bar \kappa_d| < 1.4$ at 95\% confidence level~\cite{Kagan:2014ila}.  Fixing all Higgs couplings to their SM values except for one of the up or down quark Yukawa couplings at a time instead yields $|\bar \kappa_u| < 0.98$, $|\bar \kappa_d| < 0.93$, again at 95\% confidence level~\cite{Kagan:2014ila}.  An alternative method~\cite{Zhou:2015wra} considers the inclusive $pp \to h \to 4\ell$ production rate in the off-shell region, which is unaffected by the total Higgs width; current data yields limits less sensitive by about a factor of two than the on-shell fits. 

Two completely different methods for constraining the up and down Yukawa couplings have recently been proposed.  The first relies on measuring isotope shifts in atomic clock transitions, which can be affected by Higgs exchange as well as the usual electroweak gauge boson exchange~\cite{Delaunay:2016brc}. This method depends strongly on the precision of future isotope shift measurements and on an accurate theoretical determination of the electroweak gauge contribution; nevertheless, it may yield constraints at a level comparable to the Higgs coupling fit described above.  The second relies on a future discovery of Higgs-portal dark matter; in such a scenario, if the dark matter relic density is set by the usual thermal freeze-out, current direct-detection limits already constrain the light quark Yukawa couplings at the level of $|\bar{\kappa}_{u, d}| \lesssim 0.01$~\cite{Bishara:2015cha}.

In this paper we propose a complementary technique to constrain the Higgs couplings to up and down quarks using the Higgs boson production kinematics at the LHC. 
If the shapes of the Higgs kinematic distributions from gluon-fusion production are sufficiently different from the shapes of the same distributions initiated by quark fusion, a measurement of these distributions can be used to discriminate between them and set limits on the fraction of Higgs events produced via quark fusion. There are good theoretical reasons to expect the kinematic distributions for Higgs production via $u \bar u$ or $d \bar d$ fusion to be different from those via gluon fusion.  The Higgs transverse momentum ($p_T$) distribution is shaped mainly by the additional jet radiation from the initial-state partons, which is controlled by the strong charges and spins of the initial-state partons.  Indeed, we find that the gluon-fusion process has a harder $p_T$ distribution than quark fusion, allowing these to be discriminated.  For concreteness, we parametrize the Higgs $p_T$ distributions in terms of a high-$p_T$/low-$p_T$ asymmetry parameter and determine the optimum division between the high- and low-$p_T$ regions.

We would also expect the Higgs longitudinal momentum ($p_z$) to be smaller (more central) in the gluon-fusion process and larger in the quark-fusion processes, due to the asymmetry in the average proton momentum fraction carried by a valence quark and the corresponding antiquark.  However, after taking into account the detector acceptance for the Higgs decay products in the four-lepton channel, we find that the $p_z$ distributions do not provide additional sensitivity.  This distribution may be worthy of further study in the diphoton decay channel. 

This paper is organized as follows. In Section~\ref{sec:hadron_collisions} we derive the condition on the gluon and up- and down-quark couplings which ensures that the total rate in the 4 lepton channel is the same as in the SM.  This makes our method statistically independent from the fit to signal strengths. We then compute the cross sections, branching ratios and detection efficiencies that we will need for all the relevant processes using MadGraph5\_aMC@NLO~\cite{Alwall:2014hca}. Section~\ref{sec:cfkd} defines the asymmetry observable of our method and provides sample momentum distributions from simulations. We determine the expected statistical uncertainty on the up and down quark Yukawa couplings with 300 and 3000~fb$^{-1}$ of integrated luminosity at the 13~TeV LHC.  Section~\ref{sec:conclusions1} gives some context for the strength of our constraints and summarizes our conclusions. Appendix~\ref{sec:appendix_stats} contains a derivation of the statistical error on our asymmetry observable.

\section{Higgs production cross sections}
\label{sec:hadron_collisions}

We consider $pp \to h \to 4\ell$, with $\ell = e$ or $\mu$. The background for the four-lepton final state is produced mainly by direct $Z Z^{*}$ production via quark and gluon fusion~\cite{Chatrchyan:2013mxa}. The reason for using the $4\ell$ final state is that this background is very small compared to the background in the diphoton channel, so that we can ignore it here.  We also ignore Higgs production via vector boson fusion, associated production with a $W$ or $Z$ boson, and associated production with a $t \bar t$ pair; these processes can be separated out using other kinematic features.  The observed rate for the signal process can then be written as
\begin{equation} \label{eq:R_def}
R(pp \to h \to 4\ell) = \sigma(pp \to h) \cdot \text{BR}(h \to 4 \ell) \cdot \epsilon, 
\end{equation}
where $\sigma(pp \to h)$ is the total Higgs production cross section (including only our production modes of interest), ${\rm BR}(h \to 4 \ell)$ is the branching ratio of the Higgs to the four-lepton final state, and $\epsilon$ is the detector acceptance for this final state.

Our first task is to determine the relationship between $\bar \kappa_u$, $\bar \kappa_d$, and the $hgg$ effective coupling $\kappa_g$ (defined normalized to its SM value) that must be satisfied for the $pp \to h \to 4\ell$ rate to be equal to its SM expectation.  We will assume that all other Higgs couplings besides these three are fixed to their SM values.  In addition to computing the cross sections, including interference between the processes involving $\kappa_g$ and $\bar \kappa_{u,d}$ that arises at next-to-leading order (NLO) in QCD, we must determine the detector acceptances for the four leptons for each of these processes.

\subsection{Production cross sections}

\begin{figure} 
\includegraphics{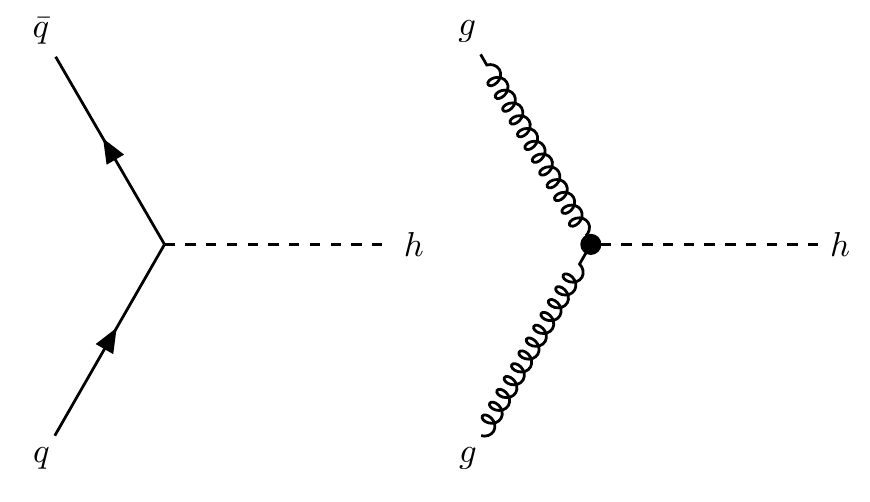}
\caption{The LO Feynman diagrams for Higgs production via quark fusion (left) and gluon fusion (right).}
\label{fig:LO_feynman}
\end{figure}

Let us first consider the cross section $\sigma(pp \to h)$. At leading order (LO),
the only two diagrams contributing to Higgs production are shown in Fig.~\ref{fig:LO_feynman}. The Higgs boson is a color singlet, so it does not couple to gluons at tree level. However, one can introduce an effective vertex (shown in Fig.~\ref{fig:LO_feynman} as a black dot) which takes into account the fact that $gg \to h$ is mediated by a heavy quark loop; this is how gluon fusion Higgs production will be handled in our Monte Carlo simulations. The two diagrams in Fig.~\ref{fig:LO_feynman} have different particles in the initial state, so they do not interfere with each other at LO.
We can then separate the gluon fusion and up- and down-quark fusion cross sections according to
\begin{eqnarray}
\sigma^{\rm LO}(pp \to h) &=& \sigma_{gg}^{\rm LO}(\kappa_g) + \sum_{q = u, d} \sigma_{q \bar q}^{\rm LO}(\bar \kappa_q) \nonumber \\
&=& \kappa_g^2 \bar \sigma_{gg}^{\rm LO} + \sum_{q = u,d} \bar \kappa_q^2 \bar \sigma_{q \bar q}^{\rm LO},
\label{eq:tot_Xs_def_LO}
\end{eqnarray}
where we define $\bar \sigma_{gg}^{\rm LO}$ as the SM (i.e., $\kappa_g = 1$) gluon fusion Higgs production cross section computed at LO, and $\bar \sigma_{u\bar u}^{\rm LO}$ and $\bar \sigma_{d \bar d}^{\rm LO}$ as the appropriate quark fusion cross sections computed at LO with $\bar \kappa_{u,d} = 1$.  Note that $\bar \sigma_{u \bar u}^{\rm LO}$ and $\bar \sigma_{d \bar d}^{\rm LO}$ are normalized in such a way that they are roughly six orders of magnitude larger than the corresponding SM cross sections.

\begin{figure} 
\includegraphics{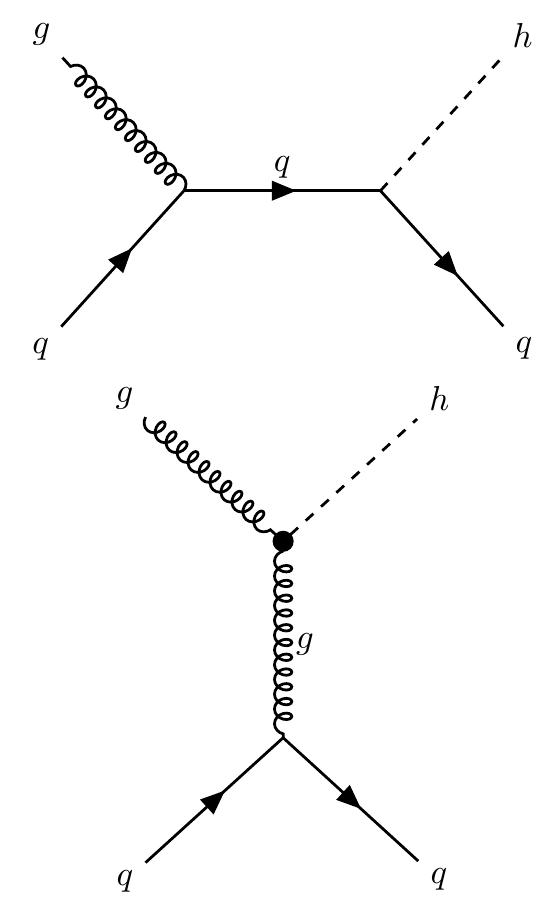}
\caption{Sample Feynman diagrams contributing to the real radiation part of the NLO cross section calculation.  These diagrams involve the same initial- and final-state particles and hence contribute to the interference term.}
\label{fig:NLO_feynman}
\end{figure}

At NLO the situation is more complicated. In addition to the virtual corrections, real radiation diagrams contribute, some of which give rise to interference as shown in Fig.~\ref{fig:NLO_feynman}. We therefore have 
\begin{eqnarray} 
\sigma^{\rm NLO}(pp \to h) &=& \sigma_{gg}^{\rm NLO}(\kappa_g) 
+ \sum_{q = u, d} \sigma_{q \bar q}^{\rm NLO}(\bar \kappa_q) \nonumber \\ 
&& + \sum_{q = u, d} \sigma_{q, {\rm int}}^{\rm NLO}(\kappa_g, \bar \kappa_q) \nonumber \\
&=& \kappa_g^2 \bar \sigma_{gg}^{\rm NLO}
+ \sum_{q = u, d} \bar \kappa_q^2 \bar \sigma_{q \bar q}^{\rm NLO} \nonumber \\
&& + \sum_{q = u, d} \kappa_g \bar \kappa_q \bar \sigma_{q, {\rm int}}^{\rm NLO},
\label{eq:Xs_in_terms_of_kappa_bar}
\end{eqnarray}
where we separate the gluon fusion, quark fusion, and interference pieces of the NLO cross section based upon their dependence on the coupling scaling factors $\kappa_g$, $\bar \kappa_u$, and $\bar \kappa_d$.  The reference cross sections $\bar \sigma_{gg}^{\rm NLO}$, $\bar \sigma_{q \bar q}^{\rm NLO}$, and $\bar \sigma_{q, {\rm int}}^{\rm NLO}$ are defined with $\kappa_g = \bar \kappa_u = \bar \kappa_d = 1$.  

In order to simulate Higgs events, we use MadGraph5\_aMC@NLO 2.2.3~\cite{Alwall:2014hca}. MadGraph automatically generates matrix elements for a process in terms of initial, final (and possibly intermediate) particles specified by the user. The default SM implementation in MadGraph explicitly sets the Yukawa couplings of the Higgs to up and down quarks equal to zero. Furthermore, one cannot easily generate $gg \to h$ since this coupling does not exist at tree level in the SM. In order to generate events which contain all the required Higgs interactions, we instead use the NLO Higgs Characterization model~\cite{Maltoni:2013sma, Demartin:2014fia, Demartin:2015uha, Demartin:2015joa}, which includes an effective vertex for the Higgs to gluon coupling as in Fig.~\ref{fig:LO_feynman}.  We introduce a further scaling factor $\kappa_g$ to modify this vertex.  We also modify the model by implementing Higgs couplings to up and down quarks, which we set equal to the Higgs coupling to bottom quarks with additional scaling factors $\bar \kappa_u$ and $\bar \kappa_d$.  We use a Higgs mass of 125~GeV throughout.

We simulate Higgs events at the 13~TeV LHC at LO and NLO in QCD, with the NLO results matched to the parton shower.  We use the NNPDF2.3\_QED (LHAPDFID = 244600) parton distribution function sets~\cite{Ball:2013hta} at the appropriate order in perturbation theory.  We shower the events using Herwig++~2.7.1~\cite{Bahr:2008pv} and cluster the jets using the anti-$k_T$ algorithm in FastJet~3.1.3~\cite{Cacciari:2011ma} (this last step is not strictly necessary for our analysis, since we will only consider the Higgs final-state momentum distributions in what follows).  In this way we obtain the reference cross sections $\bar \sigma_{gg}^{\rm NLO}$, $\bar \sigma_{q \bar q}^{\rm NLO}$, and $\bar \sigma_{q, {\rm int}}^{\rm NLO}$ as in Eq.~(\ref{eq:Xs_in_terms_of_kappa_bar}).  The interference cross sections are obtained by defining a $g,u,\bar u$ multiparticle, computing $(\bar \sigma_{gg} + \bar \sigma_{u \bar u} + \bar \sigma_{u, {\rm int}})$, and then subtracting $\bar \sigma_{gg}$ and $\bar \sigma_{u \bar u}$ (and similarly for the down quark).  Results are given in Table~\ref{tab:xsec}.  For comparison we have also computed the corresponding LO cross sections.\footnote{Note the large $k$-factor in going from LO to NLO for $\bar \sigma_{gg}$, and compare the state-of-the-art SM prediction $\bar \sigma_{gg} = 43.92$~pb from Ref.~\cite{Heinemeyer:2013tqa}.}  We do not decay the Higgs boson, so these cross sections are fully inclusive.  The uncertainties quoted in Table~\ref{tab:xsec} are the internal Monte Carlo integration uncertainties from MadGraph.
\begin{table}
\begin{tabular}{ccc}
\hline \hline
 & LO & NLO \\
\hline
$\bar \sigma_{gg}$ & $16.55 \pm 0.02$~pb & $37.3 \pm 0.3$~pb \\
$\bar \sigma_{u \bar u}$ & $13.44 \pm 0.02$~pb & $15.4 \pm 0.1$~pb \\
$\bar \sigma_{d \bar d}$ & $9.48 \pm 0.01$~pb & $11.2 \pm 0.1$~pb \\
$\bar \sigma_{u, {\rm int}}$ & -- & $14.5 \pm 0.5$~pb \\
$\bar \sigma_{d, {\rm int}}$ & -- & $10.6 \pm 0.5$~pb \\
\hline \hline
\end{tabular}
\caption{Higgs production cross sections via gluon fusion, $u \bar u$ fusion, $d \bar d$ fusion, and interference, computed using MadGraph5\_aMC@NLO 2.2.3~\cite{Alwall:2014hca} for $\kappa_g = \bar \kappa_u = \bar \kappa_d = 1$ at LO and NLO in QCD.}
\label{tab:xsec}
\end{table}

\subsection{$h \to 4 \ell$ branching ratio}

Now let us consider the branching ratio $\text{BR}(h \to 4 \ell)$. 
We would like to write this branching ratio in terms of $\kappa_g$, $\bar \kappa_u$ and $\bar \kappa_d$ as we have done with the cross section. Let $\Gamma_{\text{tot}}$ be the total width of the Higgs boson. Then, by definition, we have
\begin{equation}
\text{BR}(h \to 4 \ell) = \frac{\Gamma(h \to 4 \ell)}{\Gamma_{\text{tot}}}.
\label{eq:BR_generic_definition}
\end{equation}
The coupling modification factors $\kappa_g$, $\bar \kappa_u$ and $\bar \kappa_d$ enter through their effect on the Higgs total width.  In particular we have
\begin{eqnarray}
	\Gamma_{\rm tot} &=& \Gamma_{gg} + \sum_{q = u,d} \Gamma_{q \bar q} + \Gamma_{\rm else} \nonumber \\
    &=& \kappa_g^2 \bar \Gamma_{gg} + \sum_{q = u,d} \bar \kappa_q^2 \bar \Gamma_{q \bar q} + \Gamma_{\rm else},
    \label{eq:higgs_total_width_kappas}
\end{eqnarray}
where $\bar \Gamma_{gg}$ is the SM (i.e., $\kappa_g = 1$) Higgs decay width to two gluons, $\bar \Gamma_{u \bar u}$ and $\bar \Gamma_{d \bar d}$ are the Higgs decay widths to $u \bar u$ and $d \bar d$ respectively with $\bar \kappa_u = \bar \kappa_d = 1$, and $\Gamma_{\rm else}$ is the Higgs partial width to all other SM final states, which we hold fixed to its SM value.  The largest contribution to $\Gamma_{\rm else}$ comes from $h \to b \bar{b}$, followed by $h \to W W^{*}$. 
Because $\bar \kappa_q = 1$ implies that the $q$ quark Yukawa coupling is set equal to the bottom quark Yukawa coupling, the partial width $\bar \Gamma_{q \bar q}$ is equal to the SM Higgs partial width to $b \bar b$ up to finite bottom quark mass effects, which are at the percent level and will henceforth be neglected.  For these partial widths we will use the up-to-date SM theoretical predictions from the LHC Higgs Cross Section Working Group~\cite{Heinemeyer:2013tqa}, which are reproduced in the last column of Table~\ref{tab:widths}.  These include higher order QCD and electroweak corrections to Higgs decay partial widths, which can be quite sizable for Higgs decays to $q \bar{q}$. 
\begin{table}
\begin{tabular}{ccc}
\hline \hline
 & LO & HXSWG \\
\hline
$\bar \Gamma_{gg}$ & 0.183~MeV & 0.349~MeV \\
$\bar \Gamma_{u \bar u}$ & 4.34~MeV & 2.35~MeV \\
$\bar \Gamma_{d \bar d}$ & 4.34~MeV & 2.35~MeV \\
$\Gamma_{\rm else}$ & 5.95~MeV & 3.72~MeV \\
$\Gamma_{ZZ^*}$ & 0.090~MeV & 0.107~MeV \\
\hline \hline
\end{tabular}
\caption{Higgs partial widths for $\kappa_g = \bar \kappa_u = \bar \kappa_d = 1$.  The first column shows the LO widths computed by MadGraph and the second shows the current state-of-the-art theoretical predictions from the LHC Higgs Cross Section Working Group (HXSWG)~\cite{Heinemeyer:2013tqa}.  For the latter we take $\bar \Gamma_{u \bar u} = \bar \Gamma_{d \bar d} = \Gamma^{\rm SM}_{b \bar b}$.}
\label{tab:widths}
\end{table}

\subsection{Detector acceptance}

Finally we need to determine the detector acceptances $\epsilon$ for each of the production processes.  We compute these separately because we expect the different kinematic distributions of the different Higgs production processes to lead to different detector acceptances.  We define an acceptance for each of the NLO reference cross sections in Eq.~(\ref{eq:Xs_in_terms_of_kappa_bar}), and similarly for the LO reference cross sections.  

To compute the acceptance for, e.g., the gluon fusion process, we generate the process $gg \to h \to 4 \ell$ (including final states with electrons and/or muons), applying the following kinematic cuts at the generator level:
\begin{eqnarray}
	p_{T,\ell} &>& 10~{\rm GeV}, \nonumber \\
	|\eta_{\ell}| &<& 2.5, \nonumber \\
    |m_{4\ell} - m_h| &<& 1~{\rm GeV},
    \label{eq:cuts}
\end{eqnarray}
where $p_{T,\ell}$ is the transverse momentum and $\eta_{\ell}$ is the pseudorapidity of each of the four leptons, and $m_{4\ell}$ is the four-lepton invariant mass.  The pseudorapidity cut approximates the angular acceptance of the inner trackers of the LHC detectors.
The cut on the four-lepton invariant mass eliminates contributions from an off-shell Higgs boson, which can be significant when $m_{4\ell} > 2 M_Z$.
We then divide this decayed cross section by the corresponding reference cross section from Table~\ref{tab:xsec} and by the branching ratio for $h \to 4\ell$.  This yields the acceptance,
\begin{equation}
	\epsilon_{gg}^{\rm NLO} = \frac{\sigma^{\rm NLO}(gg \to h \to 4\ell)}{\bar \sigma_{gg}^{\rm NLO} \cdot {\rm BR}(h \to 4\ell)},
    \label{eq:efficiencydef}
\end{equation}
where we have displayed the NLO case for concreteness.  The acceptances for the interference cross sections are obtained by subtraction in a similar way as the reference cross sections.

Some comments are in order regarding the branching ratio ${\rm BR}(h \to 4\ell)$ in Eq.~(\ref{eq:efficiencydef}).  First, MadGraph always computes the branching ratios at LO, even when cross sections are being generated at NLO.  Therefore, for consistency we must divide out the LO branching ratio.  Second, the branching ratio is defined as
\begin{eqnarray}
\text{BR}(h \to 4 \ell) &=& \text{BR}(h \to Z Z^{*}) \nonumber \\
&& \hspace*{-1cm} \times [ \text{BR}(Z \to e^+ e^-) + \text{BR}(Z \to \mu^+ \mu^-)]^2,
\label{eq:BR_higgs_to_4l_in_terms_of_BRs}
\end{eqnarray}
where ${\rm BR}(Z \to e^+e^-) = {\rm BR}(Z \to \mu^+\mu^-) = 3.43 \times 10^{-2}$ as computed by MadGraph.  Here ${\rm BR}(h \to Z Z^*)$ is to be computed according to Eqs.~(\ref{eq:BR_generic_definition}) and (\ref{eq:higgs_total_width_kappas}) using the same values of $\kappa_g$, $\bar \kappa_u$ and $\bar \kappa_d$ as were used in the generation of the decayed cross section.  The relevant LO partial widths as computed by MadGraph are given in the first column of Table~\ref{tab:widths}.

The resulting detector acceptances for each of our reference cross sections are given in Table~\ref{tab:efficiencies}.  We give both LO and NLO acceptances for comparison.  Note that the NLO acceptance for the gluon fusion process is about a third lower than that at LO, but the acceptances for the $u \bar u$ and $d \bar d$ fusion processes are quite similar at LO and NLO.  At NLO, the acceptances for our reference cross sections are all roughly equal, $\epsilon_i^{\rm NLO} \sim 0.2$ to within about 20\%.
\begin{table}
\begin{tabular}{ccc}
\hline \hline
 & LO & NLO \\
\hline
$\epsilon_{gg}$ & 0.306 & 0.204 \\
$\epsilon_{u \bar u}$ & 0.184 & 0.196 \\
$\epsilon_{d \bar d}$ & 0.229 & 0.237 \\
$\epsilon_{u, {\rm int}}$ & -- & 0.186 \\
$\epsilon_{d, {\rm int}}$ & -- & 0.207 \\
\hline \hline
\end{tabular}
\caption{Detector acceptances for each of the Higgs production processes in the $4\ell$ final state.}
\label{tab:efficiencies}
\end{table}

\subsection{Signal strength constraint}

We are now in a position to extract a relationship between $\kappa_g$, $\bar \kappa_u$ and $\bar \kappa_d$ which must be satisfied for the observed Higgs signal rate in the four-lepton final state to be the same as that in the SM.  Our signal rate is given by
\begin{eqnarray}
R(pp \to h \to 4\ell) &=& \left[ \epsilon_{gg} \kappa_g^2 \bar\sigma_{gg} + \sum_{q = u,d} \epsilon_{q \bar q} \bar \kappa_q^2 \bar \sigma_{q \bar q} \right. \nonumber \\
&& \left. + \sum_{q = u,d} \epsilon_{q, {\rm int}} \kappa_g \bar \kappa_q \bar \sigma_{q, {\rm int}} \right] \nonumber \\
&& \hspace*{-0.5cm} \times \frac{\Gamma_{ZZ^*}}
{\kappa_g^2 \bar \Gamma_{gg} + \sum_{q = u,d} \bar \kappa_q^2 \bar \Gamma_{q \bar q} + \Gamma_{\rm else}}.
\end{eqnarray}
We set this equal to the SM signal rate,
\begin{equation} 
R_{\rm SM}(pp \to h \to 4 \ell) = \frac{\epsilon_{gg} \bar \sigma_{gg} \Gamma_{ZZ^*}}{\bar \Gamma_{gg} + \Gamma_{\rm else}}.
\label{eq:R_SM_definition}
\end{equation}
Rearranging this equality yields a quadratic equation for $\kappa_g$ in terms of $\bar \kappa_u$ and $\bar \kappa_d$,
\begin{equation} 
\kappa_{g}^2 + \alpha_{u} \bar{\kappa}_{u}^2 + \alpha_{d} \bar{\kappa}_{d}^2 + \beta_{u} \kappa_{g}\bar{\kappa}_{u} + \beta_{d}  \kappa_{g}\bar{\kappa}_{d} = 1,
\label{eq:SM_constraint}
\end{equation}
where the coefficients are given for $q = u,d$ by
\begin{eqnarray}
\alpha_{q} &=& \frac{\epsilon_{q \bar q}\bar{\sigma}_{q \bar{q}}(\bar \Gamma_{gg} + \Gamma_{\text{else}}) - \epsilon_{gg} \bar \sigma_{gg} \bar{\Gamma}_{q \bar{q}}}{\epsilon_{gg} \bar \sigma_{gg} \Gamma_{\text{else}}}, \nonumber \\
\beta_{q} &=& \frac{\epsilon_{q, {\rm int}} \bar \sigma_{q, {\rm int}} (\Gamma_{gg} + \Gamma_{\text{else}})}{\epsilon_{gg} \bar \sigma_{gg}\Gamma_{\text{else}}}. \label{eq:beta_definition}
\end{eqnarray}
Note that $\Gamma_{ZZ^*}$ has canceled out.  Numerical results are given in Table~\ref{tab:alphabeta}.
\begin{table}
\begin{tabular}{ccc}
\hline \hline 
 & LO & NLO \\
\hline 
$\alpha_u$ & $-0.098$ & $-0.197$ \\
$\alpha_d$ & $-0.162$ & $-0.250$ \\
$\beta_u$ & -- & $0.387$ \\
$\beta_d$ & -- & $0.315$ \\
\hline \hline 
\end{tabular}
\caption{Coefficients for the signal rate constraint in Eq.~(\ref{eq:SM_constraint}), using LO or NLO cross sections and the Higgs decay widths from the LHC Higgs Cross Section Working Group~\cite{Heinemeyer:2013tqa}.}
\label{tab:alphabeta}
\end{table}

As long as $\bar \sigma_{u, {\rm int}}$ and $\bar \sigma_{d, {\rm int}}$ are not too large, Eq.~\eqref{eq:SM_constraint} defines an ellipsoid for the allowed values of the couplings. This in itself can be used to put constraints on $\bar{\kappa}_u$ and $\bar{\kappa}_d$~\cite{Zhou:2015wra}.

\section{Kinematic discriminants}
\label{sec:cfkd}

\subsection{Asymmetry parameter}

\begin{figure}
  \includegraphics[width=\linewidth]{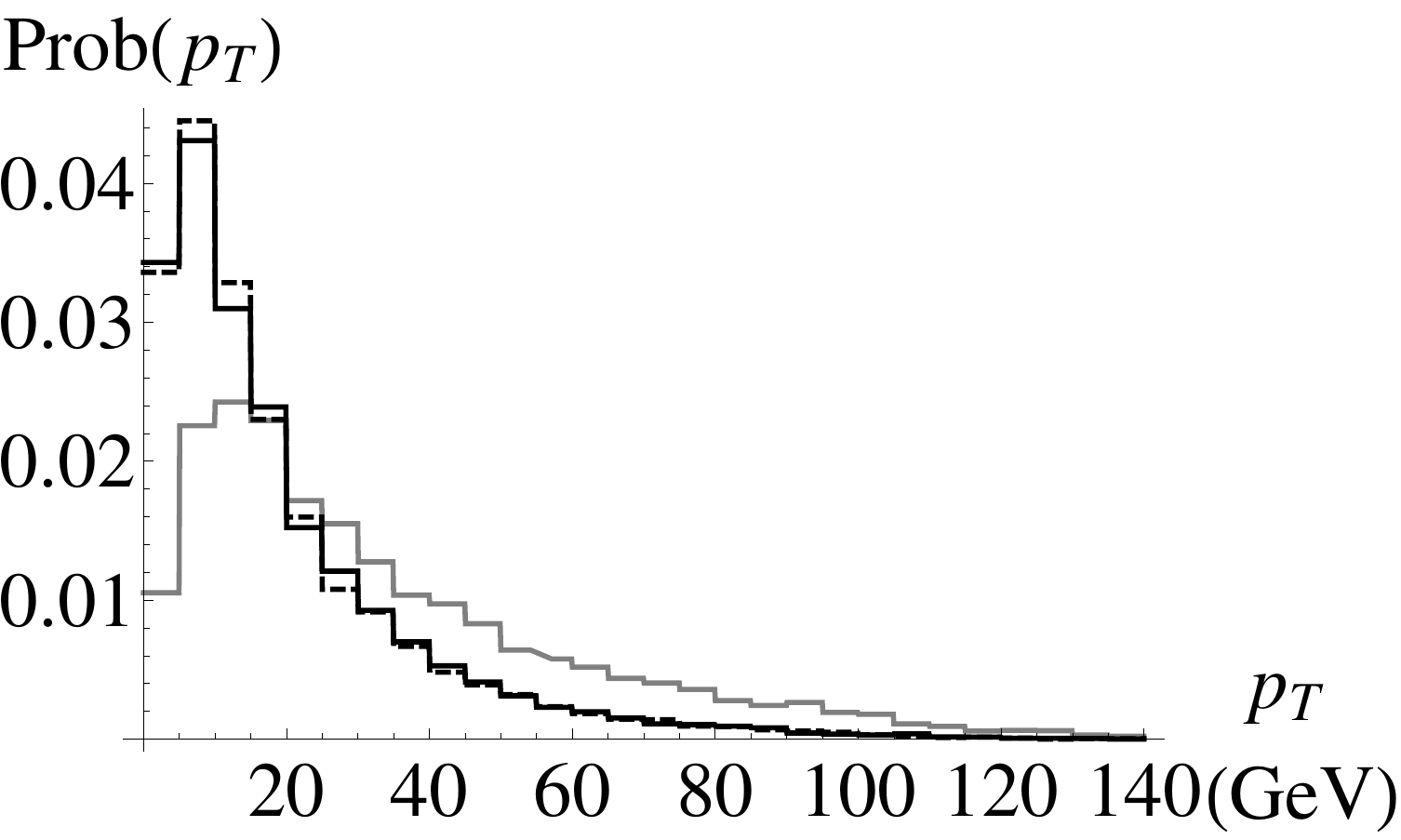} \\
  \includegraphics[width=\linewidth]{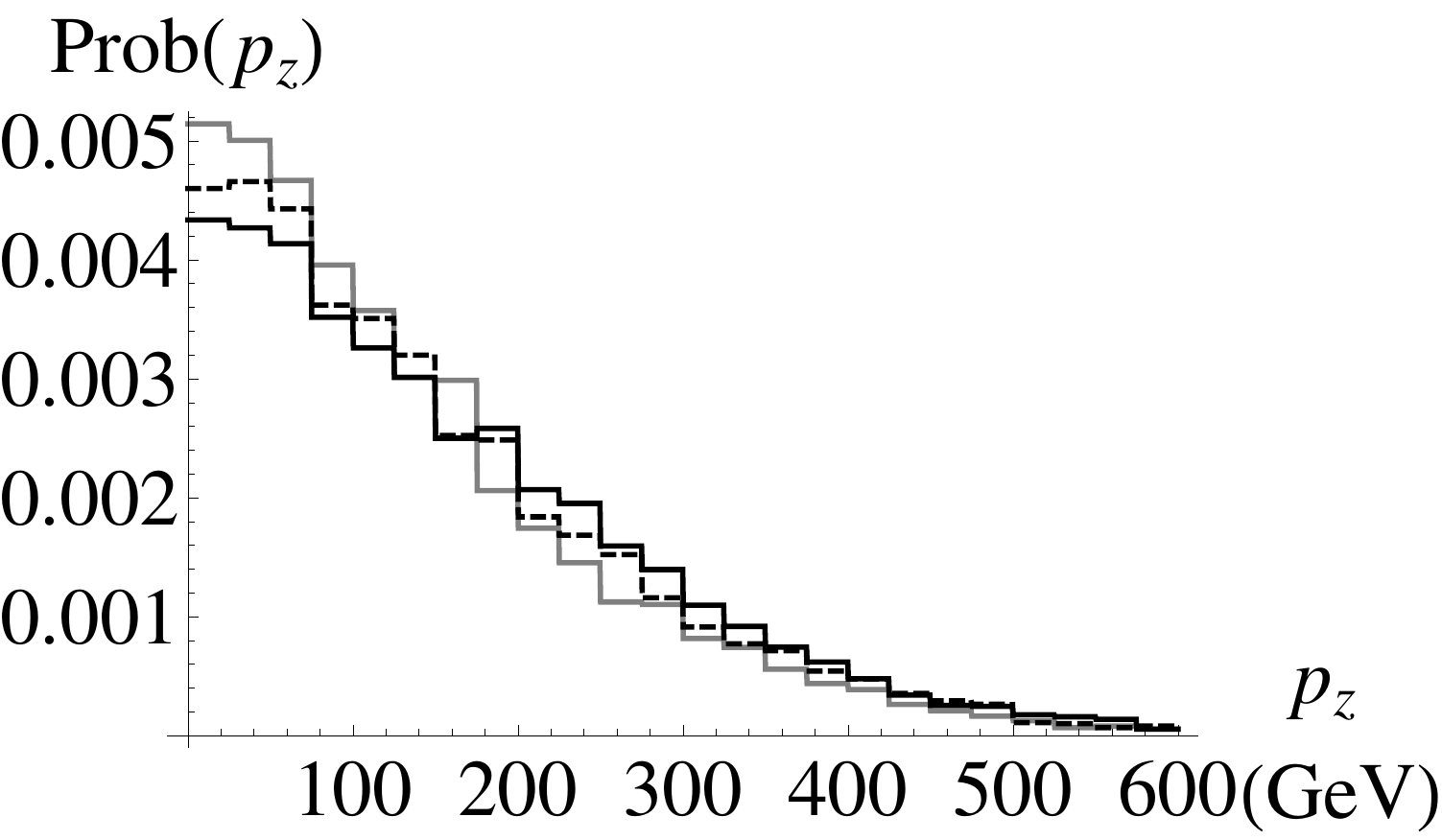}
\caption{Reconstructed Higgs $p_T$ (top) and $p_z$ (bottom) distributions for $gg \to h \to 4 \ell$ (gray), $u \bar u \to h \to 4 \ell$ (solid black), and $d \bar d \to h \to 4 \ell$ (dashed black) after cuts, from 10,000 events generated at LO in QCD.}
\label{fig:momentum_histograms_LO}
\end{figure}

\begin{figure}
  \includegraphics[width=\linewidth]{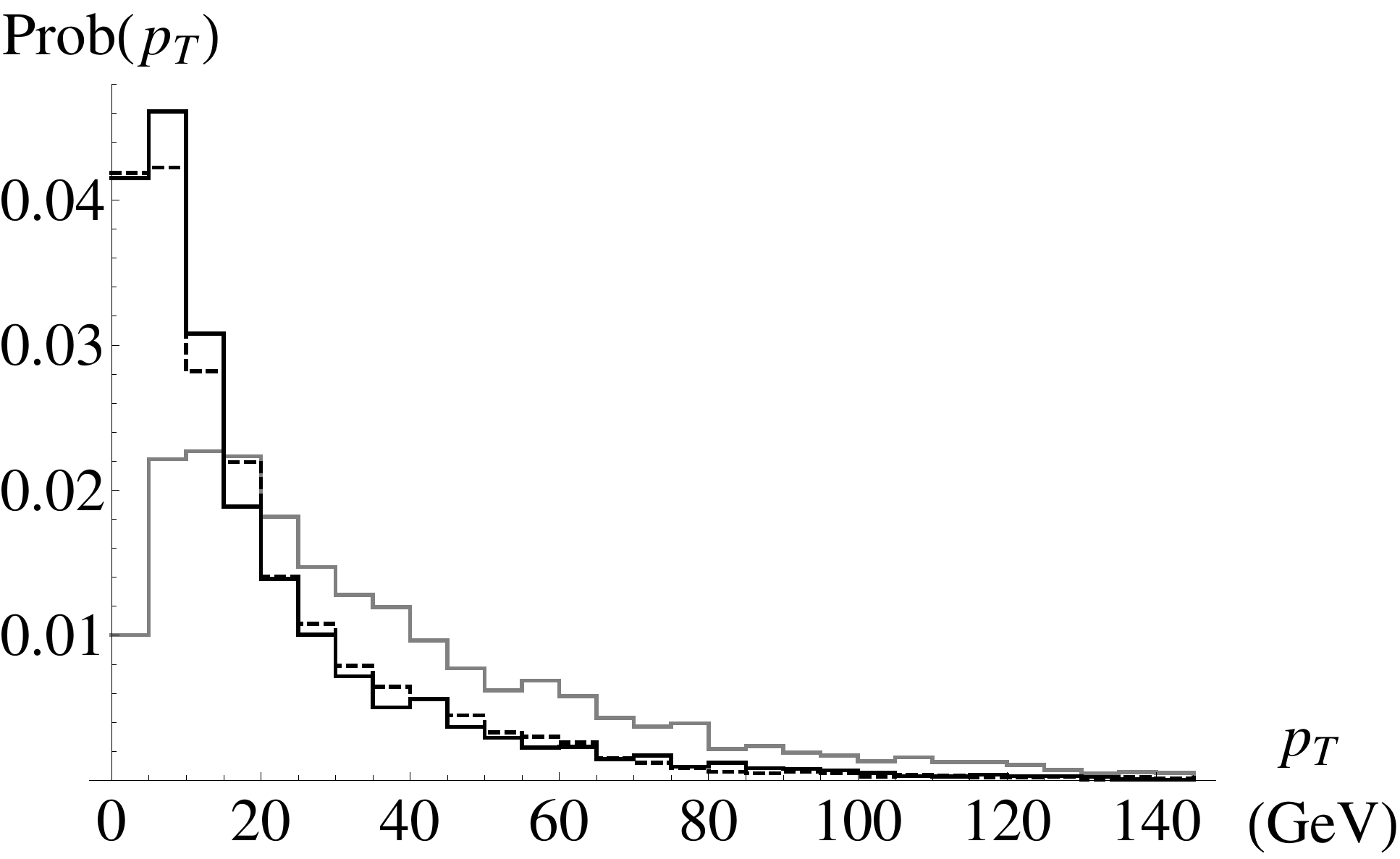} \\
  \includegraphics[width=\linewidth]{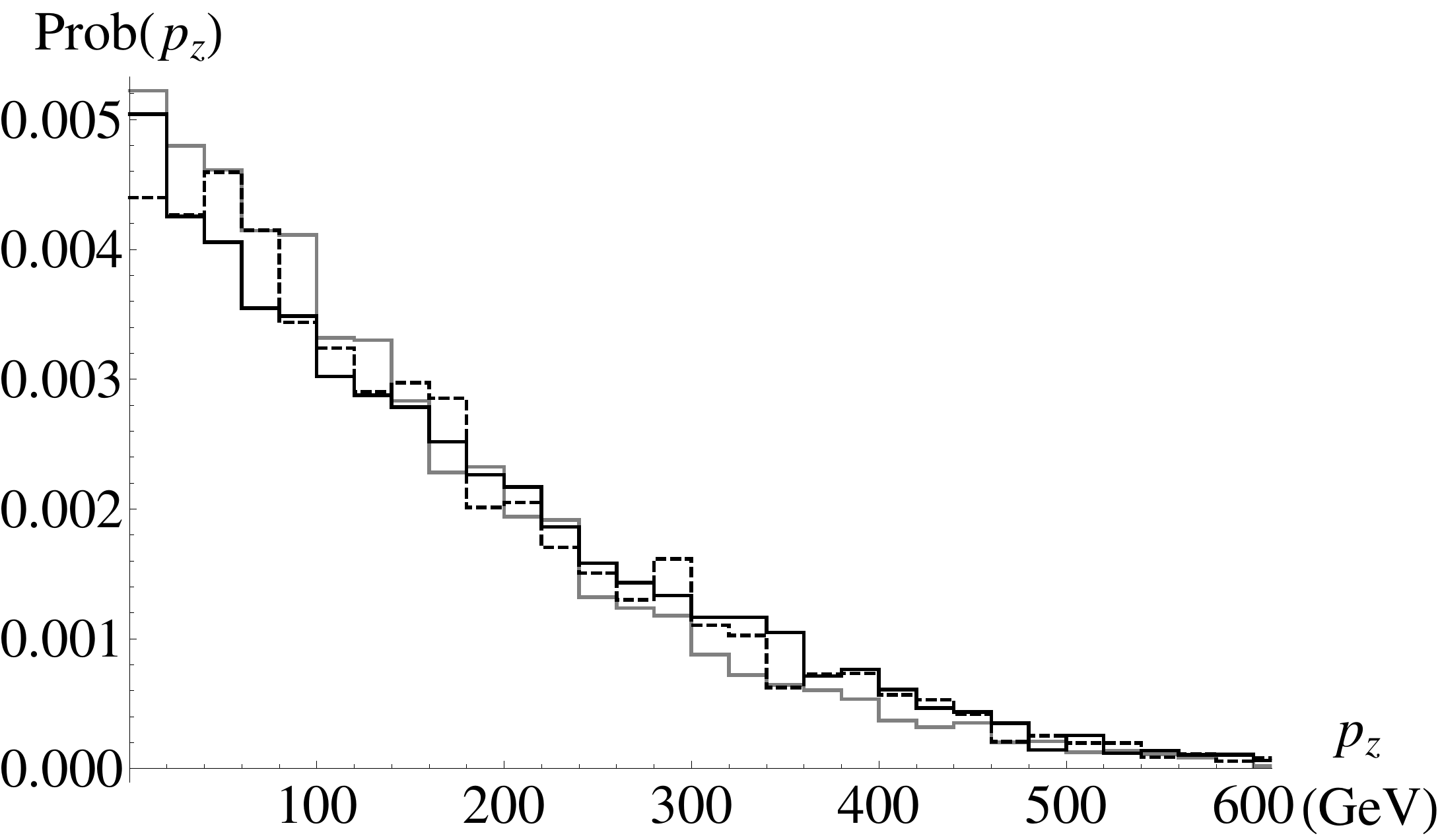}
\caption{The same as Fig.~\ref{fig:momentum_histograms_LO} but at NLO.}
\label{fig:momentum_histograms_NLO}
\end{figure}

We now turn to the Higgs kinematic distributions.
Figures~\ref{fig:momentum_histograms_LO} and~\ref{fig:momentum_histograms_NLO} show the truth-level reconstructed Higgs $p_T$ and $p_z$ for $gg \to h \to 4 \ell$, $u \bar{u} \to h \to 4 \ell$, and $d \bar{d} \to h \to 4 \ell$ at LO and NLO, respectively.  To avoid clutter, we have not plotted the interference distributions at NLO, but we do take them into account below.  At LO, the Higgs $p_T$ is due entirely to the initial-state radiation as generated by Herwig++.  The NLO calculation generates the momentum distribution of the first radiated parton at the matrix element level, so that we can expect a more accurate determination of the Higgs momentum distributions.

\begin{figure}
\centering
  \includegraphics[width=\linewidth]{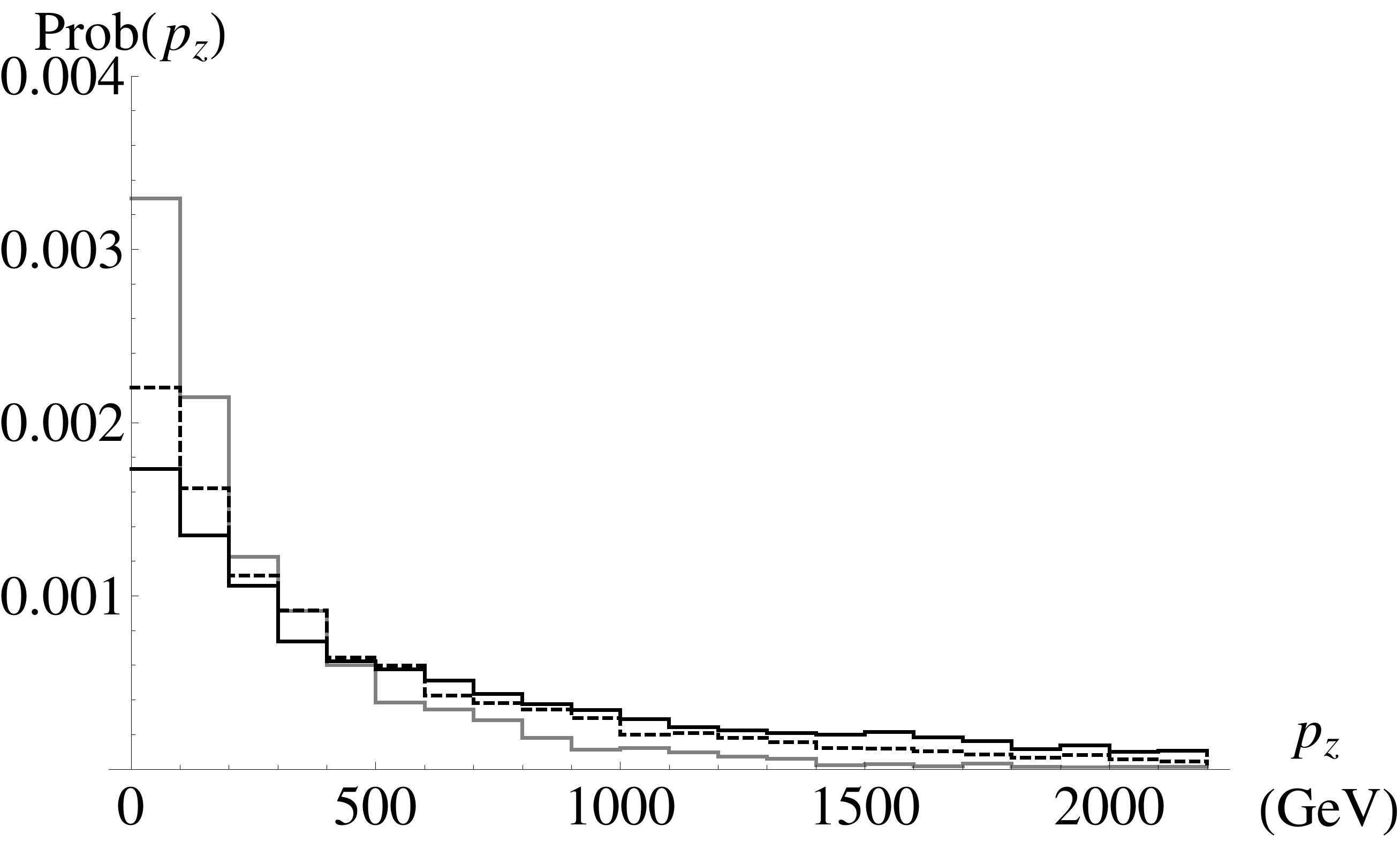}
\caption{Reconstructed Higgs $p_z$ distributions at NLO, but omitting the lepton rapidity cut in Eq.~\eqref{eq:cuts}.  The lines are the same as in Fig.~\ref{fig:momentum_histograms_NLO}.}
\label{fig:momentum_histograms_NLO_uncut}
\end{figure}

As is clear from Figs.~\ref{fig:momentum_histograms_LO} and~\ref{fig:momentum_histograms_NLO}, the Higgs $p_T$ distribution in particular is rather different for the $gg \to h \to 4\ell$ process than for the $q \bar q \to h \to 4 \ell$ processes.  This will be the basis for the discriminating power of our method.  One would also have expected the $p_z$ distribution to be different for the $gg$ fusion and $q \bar q$ fusion processes, given the very different momentum distributions carried by quarks and antiquarks in the proton.  Unfortunately, the $p_z$ distributions are made essentially identical by the lepton pseudorapidity cut, which removes the high-$p_z$ tail for Higgs production from quark fusion.  We illustrate this by showing in Fig.~\ref{fig:momentum_histograms_NLO_uncut} the Higgs $p_z$ distribution at NLO after applying only the lepton $p_T$ and $4\ell$ invariant mass cuts from Eq.~(\ref{eq:cuts}).  Indeed, we will find numerically that defining an asymmetry in a two-dimensional space of $(p_T, p_z)$ does not increase our sensitivity over using only the $p_T$ asymmetry.  The $p_z$ asymmetry may still be useful for the $h \to \gamma\gamma$ final state, in which only two objects have to fall within the pseudorapidity cut, or if the pseudorapidity coverage of the inner tracker is expanded in the course of the High-Luminosity LHC upgrades.

We define the asymmetry parameter for the reconstructed Higgs $p_T$ distribution after cuts, for each of our production processes, as 
\begin{equation} 
A^T_{j} = \frac{N(p_{T, j} > p_{T}^{\text{cut}}) - N(p_{T, j} < p_{T}^{\text{cut}})}{N_{\rm tot}},
\label{eq:A_first_definition}
\end{equation}
where $j = gg$, $u \bar u$, $d \bar d$, $u, {\rm int}$, or $d, {\rm int}$.  An analogous asymmetry can be defined for the $p_z$ distributions.  Here $p_{T}^{\text{cut}}$ is some critical momentum value around which the asymmetry parameter is calculated. The quantity $N(p_{T, j} > p_{T}^{\text{cut}})$ is the number of events of production mode $j$ with $p_T$ greater than $p_{T}^{\text{cut}}$, and $N_{\rm tot} = N(p_{T, j} > p_{T}^{\text{cut}}) + N(p_{T, j} < p_{T}^{\text{cut}})$ is the total number of events.

The asymmetry parameter measured from LHC data will be a linear combination of the asymmetry parameters for the contributing production processes, weighted by the rate for that process. Since ${\rm BR}(h \to 4\ell)$ is the same for each production process at fixed $\kappa_g$, $\bar \kappa_u$ and $\bar \kappa_d$, it cancels out of the definition in Eq.~(\ref{eq:A_first_definition}), and we can write 
\begin{widetext}
\begin{equation} 
A_T = \frac{A^T_{gg} \kappa_{g}^2 \epsilon_{gg} \bar \sigma_{gg} + \sum_{q = u, d} A^T_{q\bar{q}} \bar{\kappa}_{q}^2  \epsilon_{q\bar q} \bar{\sigma}_{q\bar{q}} + \sum_{q = u, d} A^T_{q, \text{int}} \kappa_{g} \bar{\kappa}_{q} \epsilon_{q, \text{int}} \bar{\sigma}_{q, \text{int}}}{\kappa_{g}^2 \epsilon_{gg} \bar \sigma_{gg} + \sum_{q = u, d} \bar{\kappa}_{q}^2  \epsilon_{q \bar q} \bar{\sigma}_{q\bar{q}} + \sum_{q = u, d} \kappa_{g} \bar{\kappa}_{q} \epsilon_{q, \text{int}} \bar{\sigma}_{q, \text{int}}}.
\label{eq:A_in_terms_of_kappas}
\end{equation}
\end{widetext}
An analogous expression holds for the $p_z$ asymmetry parameter.
We can eliminate $\kappa_{g}$ from Eq.~\eqref{eq:A_in_terms_of_kappas} by imposing the requirement that the total Higgs event rate in four leptons is consistent with the SM prediction, Eq.~\eqref{eq:SM_constraint}, thereby making our observable orthogonal to the total rate measurement.  We can then write $A_T$ as an analytic function of $\bar{\kappa}_{u}$ and $\bar{\kappa}_{d}$.  A measurement of $A_T$ then constrains these two parameters.

In order to implement this procedure, we must choose a value for $p_T^{\rm cut}$ and determine from Monte Carlo the asymmetry parameters $A^T_{gg}$, $A^T_{q \bar q}$, and $A^T_{q, {\rm int}}$ (with $q = u,d$).  Assuming that the measured asymmetry parameter is equal to the SM expectation, i.e., $A_T = A^T_{gg}$, we obtain the expected sensitivity to $\bar{\kappa}_{u}$ and $\bar{\kappa}_{d}$ as a function of the uncertainty on $A_T$.  

At LO, the choice of $p_T^{\rm cut}$ is straightforward: we simply maximize the difference $\Delta A^T_q \equiv A^T_{gg} - A^T_{q \bar q}$ for $q = u,d$.  Because the Higgs $p_T$ distributions are so similar for the $u \bar u$ and $d \bar d$ fusion processes (Fig.~\ref{fig:momentum_histograms_LO}), the optimum $p_T^{\rm cut}$ is the same within our Monte Carlo uncertainties for these two production processes.  We find the optimum $p_T^{\rm cut} = 18$~GeV for the LO distributions.

At NLO, the situation is more complicated due to the interference terms. Clearly we would like the resolving power to be as good as possible, which translates into the requirement that $p_T^{\text{cut}}$ should be chosen to minimize the area of the constraint contour in the $(\bar \kappa_u, \bar \kappa_d)$ plane. 
To find this optimal cut we use the heuristic procedure of computing all the asymmetries on a grid of trial $p_{T}^{\text{cut}}$ values, plotting the constraint contours for each cut, and selecting the smallest one. Using this procedure we find the optimum $p_T^{\rm cut} = 20$~GeV for the NLO distributions.  The fact that the optimal cut at NLO is so close to that found at LO gives us some confidence that the NLO corrections do not overwhelmingly change the picture. Using these cuts we compute the asymmetry parameters $A^T_{j}$ for each production process; results are given in Table~\ref{table:optimal_cuts_and_A}.
\begin{table}
\begin{tabular}{ccc}
\hline \hline 
 & LO ($p_T^{\text{cut}} = 18$~GeV) & NLO ($p_T^{\text{cut}} = 20$~GeV) \\ 
\hline
$A^T_{gg}$ & $0.29 \pm 0.01$ & $0.27 \pm 0.01$ \\ 
$A^T_{u \bar u}$ & $-0.24 \pm 0.01$ & $-0.35 \pm 0.01$ \\ 
$A^T_{d \bar d}$ & $-0.26 \pm 0.01$ & $-0.32 \pm 0.01$ \\
$A^T_{u, {\rm int}}$ & -- & $0.070 \pm 0.001$ \\ 
$A^T_{d, {\rm int}}$ & -- & $-0.014 \pm 0.001$ \\ 
\hline \hline
\end{tabular}
\caption{Asymmetry parameters $A^T_{j}$ for each production process calculated using optimized $p_T^{\rm cut}$ values at LO and NLO.  The uncertainties represent the Monte Carlo statistical uncertainties.}
\label{table:optimal_cuts_and_A}
\end{table}
%

\subsection{Sensitivity estimate}
\label{sec:constraints}

In what follows we assume that the experimental measurement of $A_T$ will be consistent with the SM expectation (i.e., $A_T = A^T_{gg}$) and proceed to estimate the constraint that can be placed upon $\bar \kappa_u$ and $\bar \kappa_d$ at the 95\% confidence level.

The statistical uncertainty on $A_T$ is given by
\begin{equation} 
\sigma_{A_T}^{\text{stat}} = \sqrt{\frac{1 - A_T^2}{N_{\text{tot}}}};
\label{eq:A_error}
\end{equation}
see Appendix~\ref{sec:appendix_stats} for a derivation.  Assuming SM production and decay, the total number of events in the four-lepton decay channel is given by
\begin{equation}
	N_{\text{tot}} = \epsilon_{gg} \bar \sigma_{gg} \text{BR}(h \to 4 \ell) \int \mathcal{L} \, dt,
\end{equation}
where $\int \mathcal{L}\,dt$ is the integrated luminosity and $\text{BR}(h \to 4 \ell) = 1.26 \times 10^{-4}$ from Ref.~\cite{Heinemeyer:2013tqa}. We give the expected number of $4 \ell$ events and the corresponding statistical uncertainty on the asymmetry parameter for various integrated luminosities in Table~\ref{table:expected_events}.
\begin{table}
\begin{tabular}{ccccc}
\hline \hline 
$\int \mathcal{L} \, dt$ & $N_{\text{tot}}$ (LO) & $\sigma_{A_T}^{\rm stat}$ (LO) & $N_{\text{tot}}$ (NLO) & $\sigma_{A_T}^{\rm stat}$ (NLO) \\ 
\hline
30~fb$^{-1}$ & $20$ & $0.22$ & $30$ & $0.18$ \\ 
300~fb$^{-1}$ & $200$ & $0.071$ & $300$ & $0.056$ \\ 
3000~fb$^{-1}$ & $2000$ & $0.022$ & $3000$ & $0.018$ \\ 
\hline \hline
\end{tabular}
\caption{Expected number of $4 \ell$ signal events for gluon-fusion Higgs production with decays to four leptons at the 13~TeV LHC, assuming SM production and decay rates, and the corresponding statistical uncertainty on the asymmetry parameter $A_T$.}
\label{table:expected_events}
\end{table}

\begin{figure}
\centering
\includegraphics[width=\linewidth]{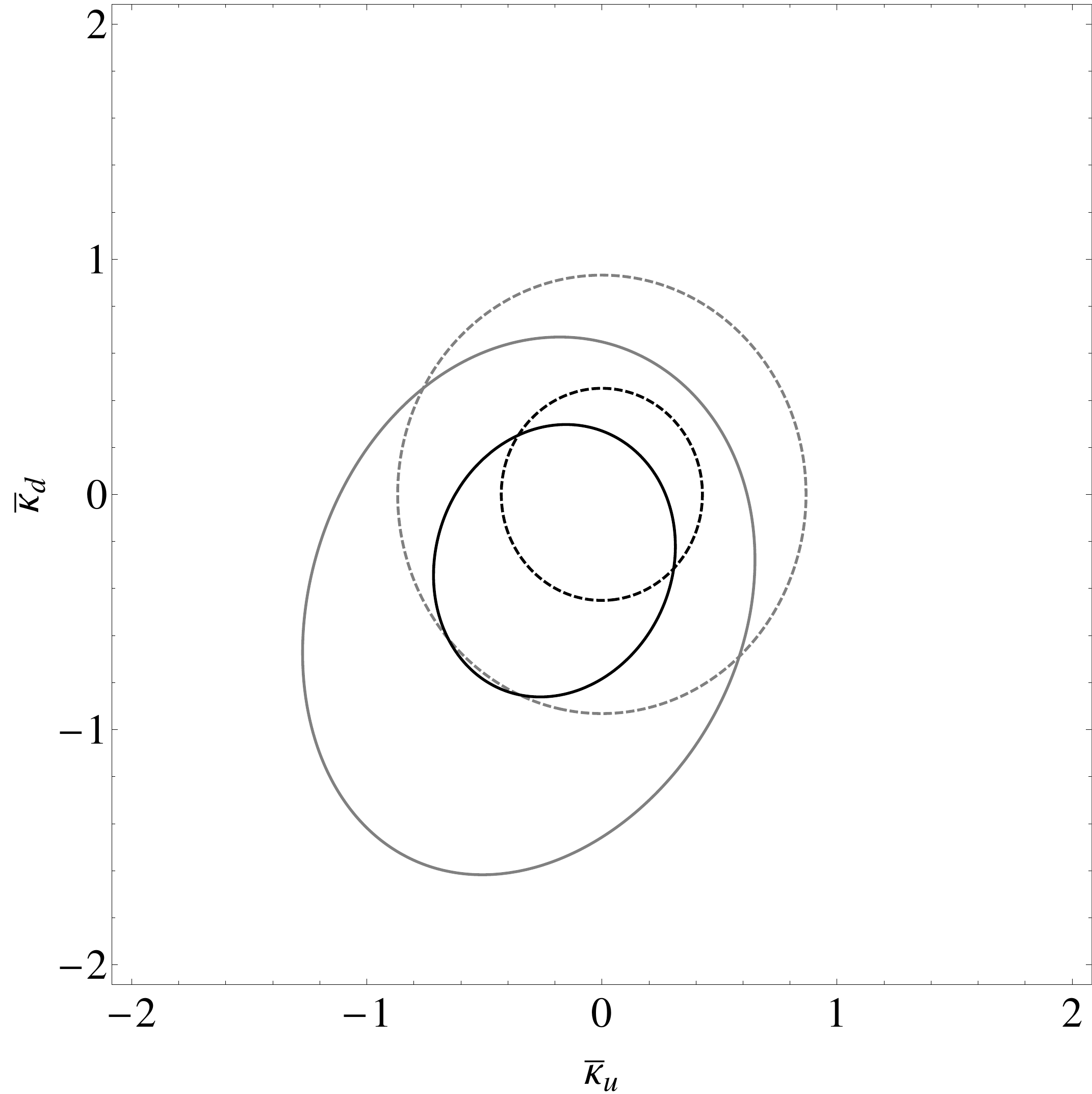}
\caption{Projected 95\% confidence level constraints on $\bar \kappa_u$ and $\bar \kappa_d$ from the Higgs $p_T$ asymmetry parameter in the four-lepton final state at LO (dashed) and NLO (solid), with 300~fb$^{-1}$ (larger gray contours) and 3000~fb$^{-1}$ (smaller black contours) at the 13~TeV LHC.  Uncertainties are statistical only.}
\label{fig:constraint_contours}
\end{figure}

Combining Eqs.~(\ref{eq:A_in_terms_of_kappas}) and~(\ref{eq:SM_constraint}), plugging in numbers, and setting the asymmetry parameter equal to its SM expectation within uncertainties, at LO we have
\begin{equation}  
A_{T}^{\text{LO}} = \frac{0.29 - 0.089 \bar{\kappa}_{u}^2 - 0.064 \bar{\kappa}_{d}^2}{1.0 + 0.59( \bar{\kappa}_{u}^2 + \bar{\kappa}_{d}^2)} = 0.29 \pm 2 \sigma_{A_T}^{\rm stat}. 
\label{eq:AT_with_numbers_LO}
\end{equation}
This expression defines a circle in the $(\bar \kappa_u, \bar \kappa_d)$ plane with radius determined by $\sigma_{A_T}^{\rm stat}$.  This is shown for 300 and 3000~fb$^{-1}$ in Fig.~\ref{fig:constraint_contours} (dashed lines).

At NLO, the functional form is more complicated; we have
\begin{widetext}
\begin{eqnarray}
\kappa_g &=& -0.19 \bar{\kappa}_u - 0.16 \bar{\kappa}_d  \pm \sqrt{1.0 + 0.23 \bar{\kappa}_u^2 + 0.27 \bar{\kappa}_d^2 + 0.061 \bar{\kappa}_u \bar{\kappa}_d}, \label{eq:kappag_NLO} \\
A_{T}^{\text{NLO}} &=& \frac{0.27 \kappa_{g}^2 - 0.14 \bar{\kappa}_u^2 - 0.11 \bar{\kappa}_d^2 + 0.025 \bar{\kappa}_u \kappa_g - 0.0040 \bar{\kappa}_d \kappa_g}{1.0 \kappa_{g}^2 + 0.40\bar{\kappa}_u^2 + 0.35 \bar{\kappa}_d^2 + 0.35 \bar{\kappa}_u \kappa_g + 0.29 \bar{\kappa}_d \kappa_g} \label{eq:AT_with_numbers_NLO}. 
\end{eqnarray}
\end{widetext}
We will choose the plus sign in the expression for $\kappa_g$, so that $\kappa_g$ is positive (choosing the minus sign is equivalent to replacing $(\bar \kappa_u, \bar \kappa_d) \to (-\bar \kappa_u, -\bar \kappa_d)$ in Fig.~\ref{fig:constraint_contours}).
The terms in $A_T^{\rm NLO}$ in which $\bar \kappa_u$ and $\bar \kappa_d$ enter linearly introduce an asymmetry in the constraint that depends on the signs of $\bar \kappa_u$ and $\bar \kappa_d$.  Setting $A_T^{\rm NLO} = 0.27 \pm 2 \sigma_{A_T}^{\rm stat}$ yields the constraint contours shown by solid lines in Fig.~\ref{fig:constraint_contours} for 300 and 3000~fb$^{-1}$.  These constraints are given numerically in Table~\ref{table:kappa_results}.
\begin{table}
\begin{tabular}{ccc}
\hline \hline 
 & 300~fb$^{-1}$ & 3000~fb$^{-1}$ \\
\hline 
$\bar \kappa_u$ & $(-1.3, 0.67)$ & $(-0.73, 0.33)$ \\
$\bar \kappa_d$ & $(-1.6, 0.69)$ & $(-0.88, 0.32)$ \\
\hline \hline
\end{tabular}
\caption{Projected 95\% confidence level constraints on $\bar \kappa_u$ and $\bar \kappa_d$ from the Higgs $p_T$ asymmetry parameter in the four-lepton final state at NLO, with 300 and 3000~fb$^{-1}$ at the 13~TeV LHC.  Uncertainties are statistical only.}
\label{table:kappa_results}
\end{table}
%

\section{Discussion and conclusions}
\label{sec:conclusions1}

To get a sense of how reasonable our results are, we calculate the individual components of the Higgs cross section and decay width for our tightest limits at 3000~fb$^{-1}$. We consider (1) $\bar \kappa_u = 0.33$, $\bar \kappa_d = 0$, for which Eq.~(\ref{eq:SM_constraint}) yields $\kappa_g = 0.949$, and (2) $\bar \kappa_d = 0.32$, $\bar \kappa_u = 0$, for which Eq.~(\ref{eq:SM_constraint}) yields $\kappa_g = 0.963$.

We first consider the Higgs production cross section.  At NLO we compute the SM Higgs production cross section from gluon fusion, $\bar \sigma_{gg} = 37.3$~pb.  For $\bar \kappa_u = 0.33$, $\bar \kappa_d = 0$, and $\kappa_g = 0.949$, we find $\sigma_{gg} = 33.6$~pb, $\sigma_{u \bar u} = 1.68$~pb, and $\sigma_{u, {\rm int}} = 4.54$~pb, for a total cross section (before cuts) of 39.8~pb.  Thus at this parameter point the $u \bar u$ production process constitutes about 4\% of the total rate and the interference term constitutes a further 11\%.

For $\bar \kappa_d = 0.32$, $\bar \kappa_u = 0$, and $\kappa_g = 0.963$, we find $\sigma_{gg} = 34.6$~pb, $\sigma_{d \bar d} = 1.15$~pb, and $\sigma_{d, {\rm int}} = 3.27$~pb, for a total cross section (before cuts) of 39.0~pb.  Thus at this parameter point the $d \bar d$ production process constitutes about 3\% of the total rate and the interference term constitutes a further 8\%.  The greater sensitivity in the $\bar \kappa_d \neq 0$ case can be explained by the greater difference between $A^T_{gg}$ and $A^T_{d, {\rm int}}$ compared to the difference between $A^T_{gg}$ and $A^T_{u, {\rm int}}$ (see Table~\ref{table:optimal_cuts_and_A}).

The Higgs branching ratios are also affected.  In the SM we have ${\rm BR}(h \to gg) = 8.6\%$~\cite{Heinemeyer:2013tqa}.  For $\bar \kappa_u = 0.33$, $\bar \kappa_d = 0$, and $\kappa_g = 0.949$, this becomes ${\rm BR}(h \to gg) = 7.3\%$ and ${\rm BR}(h \to u \bar u) = 6.0\%$.  Similarly, for $\bar \kappa_d = 0.32$, $\bar \kappa_u = 0$, and $\kappa_g = 0.963$, we obtain ${\rm BR}(h \to gg) = 7.6\%$ and ${\rm BR}(h \to d \bar d) = 5.6\%$.  If techniques to separate gluon jets from quark jets~\cite{Rentala:2013uaa} become sufficiently advanced, light quark branching fractions at this level may be able to be probed at a future International Linear $e^+e^-$ Collider.  For comparison, the SM decay branching ratio for $h \to c \bar c$ is 2.9\%~\cite{Heinemeyer:2013tqa} and for $h \to s \bar s$ is below $10^{-3}$.  

Throughout this analysis we have ignored the effect of experimental and theoretical systematic uncertainties.  These are beyond the scope of this proof-of-concept, but may be of great concern, especially the theoretical uncertainties on the Higgs $p_T$ distributions in the various production channels studied.  We note that with 3000~fb$^{-1}$, the statistical uncertainty on the asymmetry in the $4\ell$ channel is 7\%; this sets the scale for whether systematic uncertainties will have a significant effect on our results.

To summarize, we have presented a method for constraining the up and down quark Yukawa couplings at a level comparable to competing approaches using Higgs $p_T$ distributions in the four-lepton final state. Our method is orthogonal to the constraint from a global fit to Higgs signal strengths in various production and decay channels, and hence can be combined to further increase the precision. We find that 3000~fb$^{-1}$ of integrated luminosity at the 13~TeV LHC can constrain $\bar{\kappa}_{u} \lesssim 0.33$ and $\bar{\kappa}_d \lesssim 0.32$.  The constraints are weaker for negative $\bar \kappa_u$ and $\bar \kappa_d$ due to interference effects.  Including the two-photon final state may improve the sensitivity.

\vspace*{0.5cm}
{\it Note added:} While we were finalizing the manuscript, we became aware of two recent papers~\cite{Bishara:2016jga,Soreq:2016rae} that also use Higgs $p_T$ distributions to constrain the Higgs couplings to quarks.  Ref.~\cite{Bishara:2016jga} considers constraints on the bottom, charm, and strange Yukawa couplings, while Ref.~\cite{Soreq:2016rae} addresses the up and down quark Yukawa couplings.  Ref.~\cite{Soreq:2016rae} fits the Higgs $p_T$ distribution to published LHC results combining the four-lepton and two-photon final states, and extrapolates the expected sensitivity to 300~fb$^{-1}$ at 13~TeV, and find constraints on $\bar \kappa_{u,d}$ roughly comparable to ours at this luminosity.

\begin{acknowledgments}
This work was supported by the Natural Sciences and Engineering Research Council of Canada. We thank Andrea Peterson for help with MadGraph and for providing a modified version of the NLO Higgs Characterization model file with Higgs couplings to up and down quarks. 
\end{acknowledgments}

\appendix

\section{Statistical uncertainty on the asymmetry}
\label{sec:appendix_stats}

Each event that we see has a definite $p_T$ of the Higgs boson, but depending on whether or not a given value is less than $p_{T}^{\text{cut}}$ it either contributes $1$ or $0$ to the quantity $N(p_{T} < p_{T}^{\text{cut}})$. Hence, we can interpret $N(p_{T} < p_{T}^{\text{cut}})/N_{\text{tot}}$ as the sample mean of a set of $N_{\text{tot}}$ Bernoulli trials  with probability of success $\mathcal{P} = \int_{0}^{p_{T}^{\text{cut}}} f_{T}(x)\,\text{d}x$ where $f_{T}(x)$ is the underlying physical distribution of $p_{T}$. The expected value of the sample mean is the mean of the underlying Bernoulli distribution, $\mathcal{P}$. The variance of the sample mean is the the variance of the underlying Bernoulli distribution$, \mathcal{P}(1 - \mathcal{P})$, divided by the number of samples. Therefore, the expected value and variance of $A_T$ are
\begin{align}
E[A_T] &= 1 - 2 \mathcal{P} = A_T,\\
V[A_T] &= \frac{4}{N_{\text{tot}}} \mathcal{P}(1 - \mathcal{P}) = \frac{1}{N_{\text{tot}}}(1 - A_T^2) .
\end{align}
Hence, given a measurement of $A_T$, the statistical uncertainty on $A_T$ is
\begin{equation}
\sigma^{\rm stat}_{A_T} = \sqrt{\frac{1 - A_T^2}{N_{\text{tot}}}}.
\end{equation}
Why is the statistical uncertainty zero when $A_T = \pm 1$? In these cases, $p_T^{\text{cut}}$ is either at exactly zero or infinity. Hence, no matter what the distributions are doing, $A_T$ will be identically equal to $\pm 1$.


\end{document}